# Statistical theory of isotropic turbulence[1]

## Part IV: multiscales and cascade

Zheng Ran

Shanghai Institute of Applied Mathematics and Mechanics,

Shanghai University, Shanghai 200072, P.R.China

This paper is the forth part of our series of work, is devoted to the analysis on the multiscales and cascade aspects of the statistical theory of isotropic turbulence based on the new Sedov-type solution. In this paper, we use the explicit map method to analyse the nonlinear dynamical behaviour for cascade in isotorpic turbulence. This deductive scale analysis is shown to provide the first visual evidence of the celebrated Richardson cascade, and reveals in partcular its multiscale character. The results also indicate that the energy cascading process has remarkable similarities with the deterministic construction rules of the logistic map. Cascade of period-doubling bifurcations have been seen in this isotropic turbuent systems that exhibit chaotic behaviour. The ' cascade ' appears as an infinite sequence of period-doubling bifurcations.

## 1. Introduction

The cascade picture of turbulent flows takes its origin from Richardson (1922) : *…we find that convectional motions are hindered by the formation of small eddies resembling those due to dynamical instability.* Richardson cascade has played a very important role in the history of the development of turbulence, though Richardson makes no further use of this cascade picture at any time. The first concept in Richardson's view of the energy cascade is that the turbulence can be considered to be composed of eddies of different sizes. Richardson's notation is that the large eddies are unstable and break up, transferring their energy to somewhat smaller eddies. These smaller eddies undergo a similar break-up process, and transfer their energy to yet smaller eddies. This energy cascade—in which energy is transfer their energy to successively smaller and smaller eddies—continues untill the Reynolds number is sufficiently small that eddy motion is stable, and molecular viscosity is effective in dissipating the kinetic energy. Kolmogorov(1941) added to and quantified this picture, which is still the basis for nearly all work on the statistical theory of turbulence. The cascade picture is based on the intuitive notion that turbulent flows possess a hierachical structure consisting of 'eddies' (Richardson's whirls, Kolmogorov's pulsations, etc.) as a result of successive instabilities. The essence of cascade picture is in its successive hierarchical process. The general pattern of turbulent motion also can be described in the following way. The mean flow is accompanied by turbulent fluctuations imposed on it and having different scales, beginning with maximal scales of the order of the 'external scale-$L$' of turbulence to the smallest scales of the order of the distance $\eta$ at which the effect of viscosity becomes appreciable (the internal scale of turbulence). Most large scale fluctuations receive energy from the mean flow and transfer it to fluctuations of smaller scales. Thus there appears to be a flux of energy transferred

[1] The work was supported by the National Natural Science Foundation of China (Grant Nos.10272018, 10572083, 90816013)

continuously from fluctuations of large scales to those of smaller scales. Dissipation of energy, that is, transformation of energy into heat, occurs mainly in fluctuations of scale $\eta$. The amount of energy $\varepsilon$ dissipated in unit time per unit volume is the basic characteristic of turbulent motion for all scales. Richardson's hypothesis says that, in a turbulent flow, energy is continually passed down from the large-scale structures to small scales, where it is destroyed by viscous stresses. Moreover, this is a multi-stage process involving a hierarchy of vortex sizes. Kolmogorov's theory, on the other hand, asserts that the statistical properties of the small scales depend only on and on the rate at which energy is passed down the energy cascade, and most embodiments of Kolmogorov's idea of turbulent cascade rest on heuristic modelling rather than mathematical treatment of the Navier-Stokes equations. Update, it appears that these statements cannot be formally 'proven' in any deductive way.

In this paper, we will re-examine Richardson's idea of an energy cascade, in which, it is claimed, energy is passed down from large to small scales by a repeated sequence of discrete steps. We shall show that Richardson's energy cascade is a direct consequence of vortex period-doubling bifurcation, based on the new scaling equation we found before.

This paper is the forth part of our series of work, is devoted to the analysis on the multiscales and cascade aspects of the statistical theory of isotropic turbulence based on the new Sedov-type solution.

It should be pointed out that in the following sections, we will give some remarks on the subject respectively. More comprehensive accounts have been provided by Batchelor (1953), Hinze (1975), Monin and Yaglom (1975), Lesieur (1990), Frisch (1995), Pope (2000), and Davidson (2004), especially by the latest one.

## 2.The phenomenology of Richardson's cascade (Davison, 2004)

It is an empirical observation that turbulence contains a wide range of time and length sclaes. The eddies which are primarily responsible for energy transfer are the largest in the flow, and these have a size dictated by th nature of their birth. Often the large turbulent vortices arise through a distortion or instability of the mean flow vortex lines. Their size then corresponds to a length scale characteristic of mean flow, for example, the length associated with gradients in the mean velocity field.
Now the turbulence usually receives its energy from the maen flow. In a shear flow, for example, the rate of generation of turbulent energy is,

$$\rho G = \tau_{ij}^R \overline{S}_{ij} \tag{2.1}$$

Where

$$\tau_{ij}^R = -\rho \langle u_i' u_j' \rangle \tag{2.2}$$

and $\overline{S}_{ij}$ is the strain-rate of the mean flow

$$\overline{S}_{ij} = \frac{1}{2}\left(\frac{\partial \overline{u}_i}{\partial x_j} + \frac{\partial \overline{u}_j}{\partial x_i}\right) \tag{2.3}$$

Physically, this corresponds to turbulent vortices being stretched by the mean shear, increasing their energy. The eddies which are primarily responsible for this energy transfer are the largest in the flow, and these have a size dictated by the nature of their birth. Often the large turbulent vortices arise through a distortion or instability of the mean flow vortex lines. Their size then corresponds to a length scale characteristic of the mean flow, for example, the length associated with gradients in the mean velocity field.

Therefore, we have mechanical energy transferred to turblence at a large scale, and extracted at a much smaller one. The question, of course, is how does the energy get from the large scale to the small-scale structures. **Richardson attempted to bridge this gap by invoking the idea of an energy cascade. He suggested that the large structures pass their energy onto somewhat smaller ones which, in turn, pass energy onto even smaller vortices and so on.** We talk of a cascade of energy from large scale down to small. **The essential claim of Richardson is that this cascade is a multistage process, involving a hierarchy of vortices of varying size. It is conventional to talk of these different size structures as eddies, which conjures up a picture of spherical-like objects of different diameters.** However, this is a little misleading. The structures may be sheet-like or tubular in shape. It is also customary to talk of the enrgy cascade in terms of eddies continually 'breaking-up' into smaller ones as a consequence of 'instabilities'. Again, this a little misleading and is just a kind of shorthand. By break-up we really just mean that energy is being transferred from one scale to the next through a distortion of the eddy shape. Also, the use of the word instability is possibly a little inappropraate, since an 'eddy' does not represent a steady base state. The word is intended to imply that large structures can evolve into smaller ones via familiar mechanisms, some of which we might encounter in stability theory.

These statements cannot be formally 'proven' in any deductive way. The best what we can do is examine whether or not they are plausible, check that they are self-consistent, and then see how they hold uo against the experimental data.

Richardson also suggested that, at high $R_e$, viscosity plays no part in the energy cascade, except at the smallest scales. Richardson envisaged an inviscid cascade of energy down to smaller and smaller scales, the cascade being driven by inertial forces alone. The cascade is halted, however, when the structure become so small and that $R_e$ based on the small-scale eddy size is the order of unity. That is, the very smallest eddies are dissipated by viscous forces and for viscosity to be significant we need $R_e$ of order unity. In this picture, the viscous forces are paasive in nature, mopping up whatever energy cascades down from above.

Now the large-scale eddies are observed to evolve (break-up) on a timescale of $l/u$, and so on the rate at which energy passed down the cascade from sbove is

$$\Pi \propto \frac{u^3}{l}. \tag{2.4}$$

Now consider the smallest scales. Suppose they have a characteristic velocity $v$ and length scale $\eta$. Since the rate of dissipation of mechanical energy is $\nu \langle \vec{\omega} \cdot \vec{\omega} \rangle$ we have

$$\varepsilon \propto \nu \frac{v^2}{\eta^2}. \tag{2.5}$$

In homogeneous, statistically steady tubulence the rate of extraction of energy from the mean flow must equal to the rate at which the energy is passed down the energy cascade from the large scales,

$$\Pi_A \propto \frac{u^3}{l}. \tag{2.6}$$

This must also equal the rate of transfer of energy at all points in the cascade since we cannot lose or gain energy at any particular scale in a steady-on-averge flow. In particular, if $\Pi_A, \Pi_B, ...\Pi_N$ represents the energy flux at various stages the cascade thenwe have

$$\Pi_A = \Pi_B = ... = \Pi_N \propto \frac{u^3}{l}. \tag{2.7}$$

So the energy tranfer even in the small eddies is controlled by the rate of break-up of the large eddies. Finally we note that the energy flux at the end of the cascade, $\Pi_N$, must equal to the viscous dissipation rate, $\varepsilon$. In summary, then, for homogeneous, statistically staedy turbulence,

$$G = \frac{1}{\rho}\tau_{ij}^R \overline{S}_{ij} = \Pi_A = \Pi_B = ... = \Pi_N = \varepsilon. \tag{2.8}$$

Combining (2.5)-(2.6) we have

$$\eta = \left(\frac{\nu^3}{\varepsilon}\right)^{\frac{1}{4}} \tag{2.9}$$

$$v = (\nu\varepsilon)^{\frac{1}{4}} \tag{2.10}$$

These are, of course, the Kolmogorov microscales.

**Now the Richardson picture is not entirely implausible, but it does raise at least two fundamental questions. First, is there some generic process which causes energy always to pass from large to small scale and not from small to large ? Second, why must the prcoess be a multistage one ? (Davison, 2004)**

### 3. New turbulent scaling evolution equations

For complete self-preserving isotropic turbulence, the corresponding scaling equation takes the following form of Eq.(3.1). Unlike what Sedov (1944,1976,1982) has done, we can only obtain a closed equation for the length scale $l(t)$(Ran,2008).

$$\frac{d^2l}{dt^2} + \frac{(2a_1 + a_2)\nu}{2l^2}\frac{dl}{dt} - \frac{a_1 a_2 \nu^2}{2l^3} = 0 \tag{3.1}$$

This is a second class of nonlinear Lienard type equation, and one can obtain its exact solution by using the standard method. So, if we have obtained $l(t)$, by suing the second equation, one can obtain the energy of turbulence. Firstly, we are only interested in the special solutions in the following form:

$$l(t) = l_0 (t + t_0)^{\frac{1}{2}} \qquad (3.2)$$

where $l_0$ are parameters to be chosen. according to th following equation

$$l_0^4 - (2a_1 + a_2) M_0^2 + 2a_1 a_2 \nu^2 = 0 \qquad (3.3)$$

This leads

$$l_0^2 = \frac{\nu}{2} \left\{ (2a_1 + a_2) \pm \sqrt{(2a_1 - a_2)^2} \right\} \qquad (3.4)$$

Here, we would like to introduce some notations for the convenience.
If

$$l_0^2 = \frac{\nu}{2} \left\{ (2a_1 + a_2) + \sqrt{(2a_1 - a_2)^2} \right\} \qquad (3.5)$$

we denote its by P_mode.
If

$$l_0^2 = \frac{\nu}{2} \left\{ (2a_1 + a_2) - \sqrt{(2a_1 - a_2)^2} \right\} \qquad (3.6)$$

we denote its by N_mode.
But the final results will depend on the values of $\sigma$. The details of the solutions are listed in Table.1.

## Table.1 Distribution of $l_0^2$

|  | P_mode | N_mode |
| --- | --- | --- |
| $0 < \sigma \leq 1$ | $2a_1 \nu$ | $a_2 \nu$ |
| $\sigma > 1$ | $a_2 \nu$ | $2a_1 \nu$ |

These two solutions are independent on the initial condition. In this paper, we call it the first kind of similarity solution. Based on the analysis above, we know that the solutions of the turbulence scaling equation can be calssified into two different kinds: P-model and N-model, and their solutions read as follows

$$l_P = \sqrt{2a_1 \nu (t + t_0)} \qquad (3.7)$$

$$l_N = \sqrt{a_2 \nu (t + t_0)} \qquad (3.8)$$

From the small dynamics theory based on Part Ⅲ, the parameter and its variable range are given

as (also see APPENDIX A)

$$\sigma = \frac{a_2}{2a_1} \quad (3.9)$$

$$\sigma_m = \frac{2}{3} + m, \quad m = 0,1,2,...M+1 \quad (3.10)$$

A useful dimensional system must be comprised of a number of fundamental (base) entries (dimensions) that are sufficient to define the magnitude of any numerically expressible quantity. These fundamental dimensions may be chosen rather arbitrarily, but for parctical reasons, should be chosen appropriately. The seclection of a dimensional system must be carried out in two steps. The first step is to select the number of fundamental dimensions, and second is to select the standard magnititudes for these dimensions.

For a turbulent scale system, if we choose the fundamental scale as:

$$l_{ref} = l_P \equiv \sqrt{a_2 \nu (t + t_0)} \quad (3.11)$$

This analysis leads a natural description of the cascade which is a multistage process, involving a hierarchy of vortices of varying size. Here, it is conventional to talk of these different size structures as eddies[2], which conjures up a picture of spherical-like objects of different diameters, denoted as $\{l_m, m = 0,1,...,M+1\}$, and

$$l_m = \sqrt{\frac{3}{2 + 3m}} \cdot l_{ref} \quad (3.12)$$

The $l_m$ represents a hierarchy of eddy sizes from the integral scale $L$ down to $\eta$.

$$\eta < l_m < L \quad (3.13)$$

It should be noted that for all $m$, $l_m$ is a decreasing function on $m$. Then the order in volume from one generation to next is:

$$L = l_0 > l_1 > l_2 ... > l_{m-1} > l_m > l_{m+1} > .... l_M > l_{M+1} = \eta \quad (3.14)$$

Let $F_m$ be defined as follows:

$$F_m = \left(\frac{l_{ref}}{l_m}\right)^2 \quad (3.15)$$

It follows from above analysis that for each $m = 0,1,2,...,M, M+1$ straightforward manipulations on $F_m$ using the definition of Eq.(3.15) show that $F_m$ are necessarily linked by a simple recursion relation:

$$F_{m-1} - 2F_m + F_{m+1} = 0 \tag{3.16}$$

---

[2] Richardson attempted to bridge this gap by invoking the idea of an energy cascade. He suggested that the large structures pass their energy onto somewhat smaller ones which, in turn, pass energy onto even smaller vortices and so on. We talk of a cascade of energy from large scale down to small. The essential claim of Richardson is that this cascade is a multistage process, involving a hierarchy of vortices of varying size. It is conventional to talk of these different size structures as eddies, which conjures up a picture of spherical-like objects of different diameters. This remarks could be see Davidson (2004).

This relation is valid for any 'scaling system'. However, for present discussion on trubulent cascade, the whole solution of $F_m$ must satisfy two boundary conditions, where $F_0$ and $F_{M+1}$ are fixed, then

$$-2F_1 + F_2 = -F_0 \tag{3.17}$$

$$F_1 - 2F_2 + F_3 = 0 \tag{3.18}$$

$$\ldots..$$

$$F_{m-1} - 2F_m + F_{m+1} = 0 \tag{3.19}$$

$$\ldots..$$

$$F_{M-1} - 2F_M = -F_{M+1} \tag{3.20}$$

These two boundary conditions uniquely determine the solution, and the final expression of $F_m$ is then

$$F_m = \frac{M+1-m}{M+1} \cdot F_0 + \frac{m}{M+1} \cdot F_{M+1} \tag{3.21}$$

The substituting of the defintition of $F_m$ into eq. (3.21) leads:

$$\frac{1}{l_m^2} = \frac{M+1-m}{M+1} \cdot \frac{1}{L^2} + \frac{m}{M+1} \cdot \frac{1}{\eta^2} \tag{3.22}$$

Moreover, we have

$$\frac{1}{l_{m+1}^2} = \frac{M+1-(m+1)}{M+1} \cdot \frac{1}{L^2} + \frac{m+1}{M+1} \cdot \frac{1}{\eta^2} \tag{3.23}$$

It is natural to ask what their interpretaion is in the present framework for $l_m$ and $l_{m+1}$. We would like to argue that

$$\frac{1}{l_{m+1}^2} - \frac{1}{l_m^2} = \frac{1}{M+1}\left[\frac{1}{\eta^2} - \frac{1}{L^2}\right] \tag{3.24}$$

It follows that (3.24) may be rewitten in the symbolic form

$$l_{m+1}^2 = \frac{l_m^2}{1 + A \cdot l_m^2} \tag{3.25}$$

where

$$A \equiv \frac{1}{M+1}\left[\frac{1}{\eta^2} - \frac{1}{L^2}\right]. \tag{3.26}$$

Accordingly, if the fundamental scale is choose as follows

$$l_R = \sqrt{2a_1 \nu(t+t_0)} \tag{3.27}$$

It is conventional to talk of these different size structures as eddies, which conjures up a picture of spherical-like objects of different diameters. Here we denote as $\{q_m, m=0,1,2,...,M+1\}$.

$$q_m = \sqrt{a_2 \nu(t+t_0)} = \sqrt{\sigma_m} l_R \tag{3.28}$$

Let $G_m$ be defined as follows:

$$G_m = \left(\frac{q_m}{l_R}\right)^2 \tag{3.29}$$

It follows from above analysis that for each $m=0,1,2,...,M,M+1$ straightforward manipulations on $G_m$ using the definition of Eq.(3.29) show that $G_m$ are necessarily linked by a simple recusion relation:

$$G_{m-1} - 2G_m + G_{m+1} = 0 \tag{3.30}$$

From eq.(3.28), we know that for all $m$, $q_m$ is an increasing function on $m$, hence, the consistent boundary condition nust be chosen as follows:

$$q_0 = \eta \tag{3.31a}$$

$$q_{M+1} = L \tag{3.31b}$$

Based on above analysis, we have the final expression for $q_m$

$$q_m^2 = \frac{M+1-m}{M+1}\eta^2 + \frac{m}{M+1}L^2 \tag{3.32}$$

Hnce, we have the recursion equation

$$q_{m+1}^2 - q_m^2 = \frac{1}{M+1} \cdot L^2 - \frac{1}{M+1} \cdot \eta^2 \tag{3.33}$$

At last we have another expression for the scale

$$q_{m+1}^2 = q_m^2 + B \tag{3.34}$$

where

$$B \equiv \frac{1}{M+1} \cdot L^2 - \frac{1}{M+1} \cdot \eta^2 \tag{3.35}$$

## 4. Logistic map for isotropic turbulence

In section 3, we have obtained two different expressions of the turbulent scales. Physically, the ture values of the turbulent scale area will be the arithmetic mean of these two different scales areas, here, we have

$$y_m \equiv \frac{l_m^2 + q_m^2}{2} \tag{4.1}$$

In analogy to the $y_m$ case, we can produce the solution in the form

$$y_{m+1} = \frac{l_{m+1}^2 + q_{m+1}^2}{2} \tag{4.2}$$

The replace of $l_{m+1}$, $q_{m+1}$ to deduce:

$$2y_{m+1} = \frac{1}{\frac{1}{l_m^2} + A} + q_m^2 + B \tag{4.3}$$

If

$$A \ll 1 \tag{4.4}$$

We have:

$$\frac{1}{1+Al_m^2} \approx 1 - Al_m^2 \tag{4.5}$$

The substituting leads

$$2y_{m+1} = l_m^2 \cdot (1 - Al_m^2) + q_m^2 + B \tag{4.6}$$

Indeed, we see that

$$l_m^2 = 2y_m - q_m^2 \tag{4.7}$$

The final expression of $y_{m+1}$ could be written as

$$\begin{aligned}
2y_{m+1} &= 2y_m - q_m^2 - A \cdot (2y_m - q_m^2)^2 + q_m^2 + B \\
&= -4A \cdot y_m^2 + (2 + 2Aq_m^2)y_m + (B - Aq_m^4)
\end{aligned} \tag{4.8}$$

We can convert this into a differnce equation only for $y_m$, in view of the assumpition of (4.4):

$$y_{m+1} = -2Ay_m^2 + y_m + \frac{1}{2}B \tag{4.9}$$

The map (4.9) is an instance of logistic map. Here, we could as well call it the logistic map of isotropic turbulence.

The quadratic law (4.9) which we have explored so far is just one of a universe of feedback system which display very complicated behavior. The expression $x^2 + c$ is another example. If we carried out experiments analogous to that of (4.9) for $c = -2$, we would observe exactly the same behavior. The reason is simply that two quadratic processes can be identified by means of a coordinate transformation, i.e., they really are the same.

Why should we look at $x^2 + c$ when the dynamics for iterators to this formula are the same (up to same coordinate transformation) as for (4.9)? There are many different problems to be solved with quadratic iterations, and indeed, in principle it does not matter which quadratic is taken because all are equivalent. However, the mathematical formulation of these problems and their solutions will be more illuminating ( and perhaps less complex ) depending on the particular quadratic we pick. Therefore, in each case, we may choose the quadratic transformation which suits best the problem on hand.

We will restrict ourselves to the iteration

$$z_{m+1} = g(z_m; c) \tag{4.10}$$

$$g(z; c) = z^2 + c \tag{4.11}$$

where

$$c = \frac{1}{4} - AB \tag{4.12}$$

(The details could be seen in the **APPENDIX B**).

At this stage, we may say that the key problem we faced up is how to derive the bifurcation parameter $c$ out, under the definition by eq.(4.12).

Based on the former analysis, we have

$$AB = \frac{1}{(M+1)^2} \cdot \left[L^2 - \eta^2\right] \cdot \left[\frac{1}{\eta^2} - \frac{1}{L^2}\right]. \tag{4.13}$$

Hence,

$$AB = \frac{1}{(M+1)^2} \cdot \left[\left(\frac{L}{\eta}\right)^2 - 2 + \left(\frac{\eta}{L}\right)^2\right] \tag{4.14}$$

We note that : the ratio of the smallest to largest scales are readily determined from the definition of the Kolmogorov scales. The results are[3]

$$\frac{L}{\eta} = \alpha \cdot \left(\frac{R}{R_c}\right)^{\frac{3}{2}} \tag{4.15}$$

In the meantime, let us set

$$(M+1)^2 = \beta \cdot \left(\frac{R}{R_c}\right)^a \tag{4.16}$$

---

[3] We note that similar consideration have enable Landau and Lifshitz to estimate the number of degree of freedom in a developed turbulent flow. In general, $N \propto R^{\frac{9}{4}}$ or, more precisely, $N \propto \left(\frac{R}{R_c}\right)^{\frac{9}{4}}$. The numerical factor $R_c$ must be included, since $N$ is of the order of unity when $R \approx R_c$ and not when $R_e \approx 1$.

were $\alpha, \beta$ are constants we must choose. The substituting of eq.(4.15) and (4.16) into (4.14) lead

$$c = \frac{1}{4} - \frac{1}{\beta \bar{R}^a} \cdot \left[ \alpha \cdot \bar{R}^{\frac{3}{2}} - 2 + \frac{1}{\alpha} \cdot \bar{R}^{-\frac{3}{2}} \right] \tag{4.17}$$

where $\bar{R} \equiv \frac{R}{R_c}$.

In order to deduce precise consequences from them, it is worthwhile to provide here more precise statements of chaos theory on the range of the bifurcation parameter $c$. According to chaos theory, we know that : for the quadratic form written by eq. (4.11), we restrict the control paprameter $c$ to the range

$$-2 \leq c \leq \frac{1}{4} \tag{4.18}$$

So that (4.11) maps the interval $0 \leq z \leq 1$ into itself.

To find $a, \alpha, \beta$, we need to prvide two boundary conditions according to the range of the control paramete, we outline the algrebra involved the rest of the solution as follows:

(1) $\bar{R} = \infty, c_\infty = -2$;

Comparing the coefficients, we have

$$\frac{3}{2} - a = 0 \tag{4.19}$$

$$\frac{1}{4} - \frac{\alpha}{\beta} = -2 \tag{4.20}$$

(2) $\bar{R} = 1, \ c = \frac{1}{4}$.

Hence, we have

$$\alpha - 2 + \frac{1}{\alpha} = 0 \tag{4.21}$$

$$\alpha = 1 \tag{4.22}$$

Based on our results so far, we have the final expression of the bifurcation parameter for isotropic turbulence:

$$c = -2 + \frac{9}{2} \cdot \left( \frac{R}{R_c} \right)^{-\frac{3}{2}} - \frac{9}{4} \cdot \left( \frac{R}{R_c} \right)^{-3} \tag{4.23}$$

Note that the control parameter $c$ only depends on the relative Reynolds number.

In the further investigation, in fact, we would not use the orginal form of eq. (4.11). Moreover, we would like use the standard form[4] which is used in the most chaos theory

---

[4] See : May, R.M., Simple mathematical models with very complicated dynamics. Nature, vol.261,459 (1976).

$$x_{m+1} = ax_m(1 - x_m) \qquad (4.24)$$

It is easy to show that

$$a = 1 + \sqrt{1 - 4c} \qquad (4.25)$$

Note that the control parameter $a$ only depends on the relative Reynolds number. $x_m$ could be regarded as the mean area of the Taylor microscale in isotropic turbulence.

Our above discussion of chaos equation opens a door to a very interesting topic. It is well known that the chaotic behaviour of solution of the logistic equation is a direct result of non-linearity. When the non-linear term is relatviely weak the solutions are well behaved. However, as the relatively magnitude of the non-linear term is increased the solutions become increasingly complex, passing through a sequence of bifurcation, each bifurcation leading to a more complex state. Eventually the solutions become so intricate and complex that they are, to all intent and purposes, unpredicatble. In short, the solutions are chaotic.

## 5. Period doubling bifurcation and turbulent cascade

Chaos theory began at the end of last centure with some great initial ideas, concepts and results of monumental French mathematician Heri Poincare. Also the more recent path of the theory has many fascinating success stories. Probably the most beautiful and important one is the theme of this section. It is known as the route from order to chaos, or Feigenbaum's universality. One of the great surprises revealed through the studies of the quadratic iterator (4.24) is that both antagonistic state can be ruled by a single law. An even bigger surprise was the discovery that there is a very well defined ' route' which leads from one state-order-into the other state-chaos. Furthermore, it was recognized that this route is universal. Here, 'route' means that there are abrupt qualitative change—called bifurcations—which make the transition from order into chaos like a schedule, and ' universal ' means that these bifurcations can be found in many natural systems both qualitatively and quantitatively.

The following computer experiment turns out to be loaded with marvellous scientific discoveries. Here is the experiment. We want to explore the behaviour of the quadratic iterator (4.24) for all values of the parameter $a$ between 1 to 4. Figrue 1. shows that the results for all parameter $a$. We note that for parameter $a > 3$ the final state is not a mere point but a collection of 2,4 or more points. For $a = 4$, of course we have the chaos, and the points of the final state densely fill up the complete interval. Sometimes this image is also called the Feigenbaum diagram.Indeed, this diagram is a remarkable fractal, and later we will see that it is closely related to the famous Mandelbrot set.

One essential structure seen in the Feigenbaum diagram 1 is that of a branching tree which portrays the qualitative changes in the dynamical behaviour of the iterator (4.24). Out of a major stem we see two branches bifurcating and out of these branches we see two branches bifurcating again, and then two branches bifurcating out of these again, and so on. This is period-doubling regime of the scenario.

Let us explain very cludely what period-doubling means. Where we see just one branch the long-term behaviour of the system tends towards a fixed final state, which, hwever, depends on the parameter $a$. This final state will be reached no matter where- at which initial state we start. When we see two branches this just means that the long- term behaviour of the system is now alternating between two different states now, a lower one and an upper one. This is called periodic behaviour. Since there are two states now, we say that the period is two. Now, when we see four branches all that has happened is that the period-doubling :

$$1 \rightarrow 2^1 \rightarrow 2^2 \rightarrow 2^3 \rightarrow 2^4 \rightarrow .....$$

Beyond this period-doubling cascade at the right end of the figure we see a structure with a lot of detailed and remarkable designs. Chaos has set in, and eventually, at $a = 4$, chaos governs the whole interval from 0 to 1.

The Feigenbaum diagram has feartures that are both of a qualitatvie nautre and a quantitative one. The qualitative features are best analyzed through the methodology of fractal geometry. The strcture in figure 2. has self-similarity properties, which, we will now show, means that the route from order to chaos is one with infinite detail and complexity. In other wors, the final state diagram is a self-similarity structure.

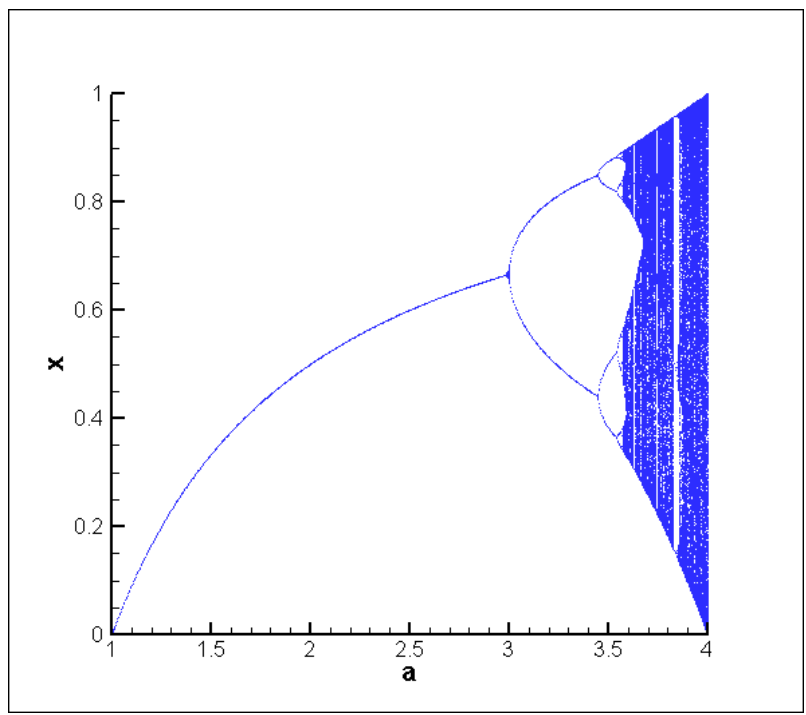

**Figure 1. The Feigenbaum diagram of isotropic turbulence**

Cascade of period-doubling bifurcations have been seen in the great majority of low-dimensional systems that exhibit chaotic behaviour. A ' cascade ' appears as an infinite sequence of period-doubling bifurcations. A stable periodic orbit is seen become unstable as a parameter is increased or decreased and is replaced by a stable periodic orbit of twice its period. This orbit in turn becomes unstable and is replaced by a new stable orbit with its period again doubled, and the process continues through an infinity of such period-doubling bifurcations.

It is well-known that the logistic equation was originally proposed for the description of the dynamics of a population of organisms that appear in discrete generations, such as insects. May[5] present a number of examples ranging from genetic problems to sociology which are modelled by equation of the type (4.24). The logistic equation is one of the most often discussed prototypes of the complex behavior of deterministic systems. The best visible is a sequence of successive period doublings. It seems that periodic points first appear in order $1, 2, 4, 8, \ldots, 2^n, \ldots$ occur as $a$ increases. Specifically, let $\dfrac{R_m}{R_c}$ denote the value of bifurcation where a $2^n$-cycle first appears. Then table.1 reveals that the critical Reynolds numbers bifurcation values of further period-doubling. Note that the successive bifurcations come faster and faster. Ultimately the bifurcation parameter converges to a limiting value $a_\infty$. For $a > a_\infty$, the orbit diagram reveals an unexpected mixture of order and chaos, with periodic windows interspersed between chaotic clouds of dots. This implies the transition from laminar to turbulence.

**TABLE.1 The critical Reynolds number for period doubling bifurcation**

| $m$ | | $\dfrac{R_m}{R_c}$ |
|---|---|---|
| 1 | $1 \to 2$ | 2.080083823 |
| 2 | $2 \to 2^2$ | 3.096733934 |
| 3 | $2^2 \to 2^3$ | 3.511518701 |
| 4 | $2^3 \to 2^4$ | 3.619876484 |
| 5 | $2^4 \to 2^5$ | 3.644190644 |
| 6 | $2^5 \to 2^6$ | 3.649451759 |
| … | … | … |
| $\infty$ | $2^\infty \to$ chaos | 3.650890633 |

---

[5] See : May, R.M., Simple mathematical models with very complicated dynamics. Nature, vol.261, 459 (1976).

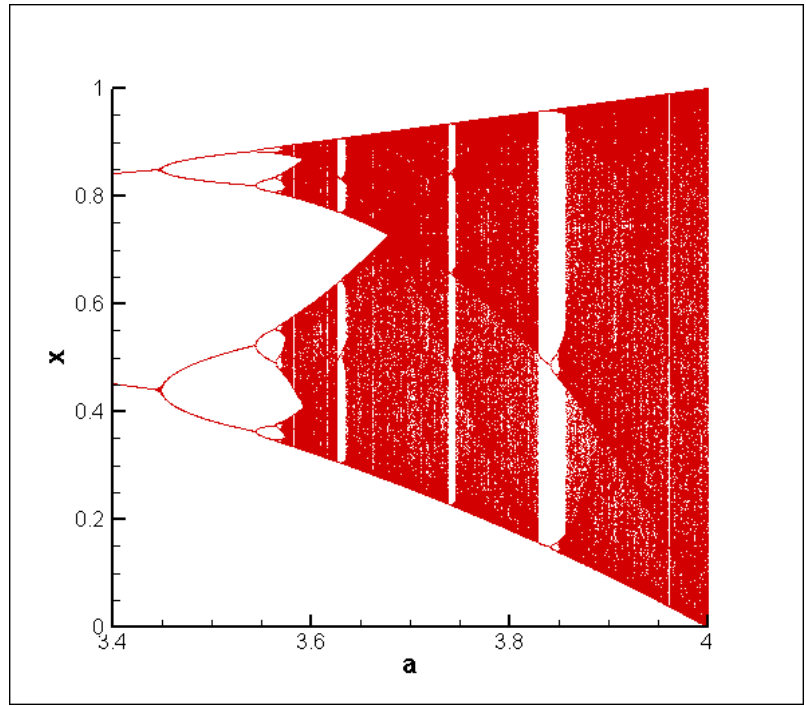

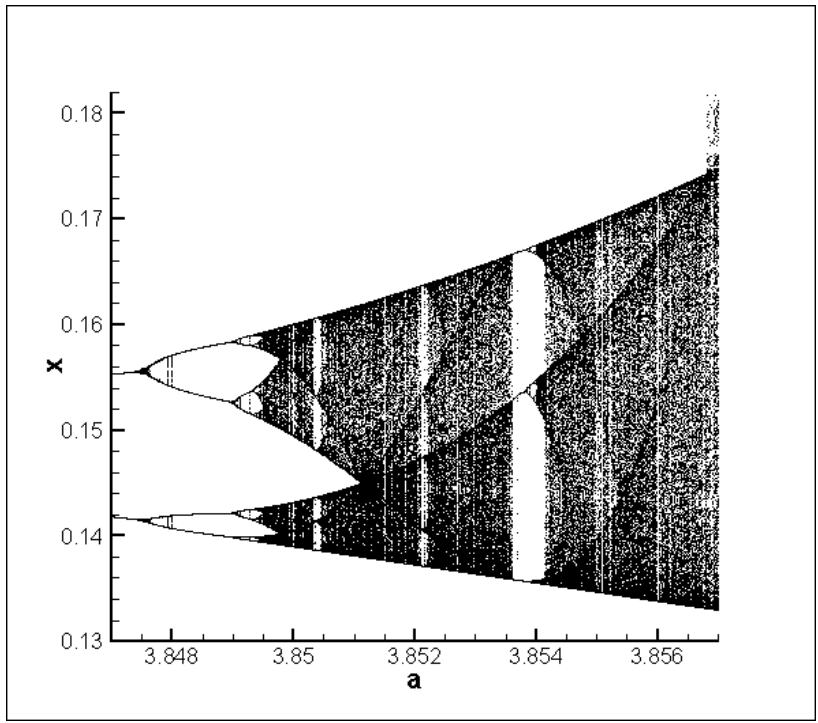

**Figure 2. The self-similarity structure in Feigenbaum diagram of isotropic turbulence**

## 6. Conclusions

One of the main goals in the development of theory of chaotic dynamical system has been to make progress in understanding of turbulence. The attempts to related turbulence to chaotic motion got strong impetus from the celebrated paper by Ruelle and Takens . Considerable success has been achieved mainly in the area: the onset of turbulence. For fully developed turbulence, many questions remain unanswered. The aim of this paper is to show that there are dynamical systems that are much simpler than the Navier-Stokes equations but that can still have turbulent states and for which many concepts developed in the theory of dynamical systems can be successfully applied. In this connection we advocate a broader use of the universal properties of a wide range of isotropic turbulence phenomena. Even for the case of fully developed turbulence, which contains an extreme range of relevant length scales, it is possible, by using the present model, to reproduce a surprising variety of relevant features, such as multifractal cascade, intermittency.

The work was supported by the National Natural Science Foundation of China (Grant Nos.10272018, 10572083).

# APPENDIX A: Equivalence of quadratic polynomials

## A.1 Spectra based on the Sedov-type solution

For isotropic turbulence, the Karman-Howarth equation, which stems from the Navier-Stokes equations, fully describes the dynamics of the two-point velocity correlation. It does not, however, provide a very clear picture of the processes involved in the energy cascade. Some further insights can be gained by examining the Navier-Stokes equations in the wave-numbers space. In this section, we will examine the energy spectrum of isotropic turbulence based on the exact solution. Several functional forms for the energy spectrum in small wave range, inertial-range have been proposed based on the asymptotic analysis.

In the following analysis, we introduce two alternative parameters denoted by $a_1, \sigma$, while the Sedov-type solution could be rewritten as

$$f(\xi) = e^{-\frac{a_1}{4}\xi^2} {}_1F_1\left(\frac{5}{2} - \sigma, \frac{5}{2}, \frac{a_1}{4}\xi^2\right) \tag{A.1.1}$$

or

$$f(\xi) = {}_1F_1\left(\sigma, \frac{5}{2}, -\frac{a_1}{4}\xi^2\right) \tag{A.1.2}$$

One-dimensional energy spectra could be deduced directly from this solution.

A turbulent flow varies randomly in all three space direction and in time. Experimental measurements, say of velocity, may be made along a straight line at a fixed time, at a fixed position as function of time, or following a moving fluid point as function of time. A measurement of this kind generates a random function of position or time. If the function is stationary or homogeneous, an autocorrelation can be formed and a spectrum can be computed. If the autocorrelation is a function of a time interval, the transform variable is wave number. Spectra obtained in this way are called one-dimensional spectra because the measurements production them were taken in one dimension. The one-dimensional spectra that are most often measured are the one-dimensional Fourier transforms of a longitudinal or transverse correlation. There is no uniformity of notation for one-dimensional energy spectra. Batchelor (1953), Hinze (1975), Tennekes and Lumley (1972), and Monin and Yaglom (1975) all use different symbols. We shall adopt the same convention as Tennekes and Lumley for the one-dimensional spectra of $u^2 f$ and $u^2 g$, that is $F_{11}$ and $F_{22}$. (Davidson,2004)

They appear particularly in experimental papers as the quantities most commonly measured in experiment. They are

$$F_{11}(k,t) = \frac{1}{\pi}\int_0^\infty u^2 f(r,t)\cos(kr)dr \tag{A.1.3}$$

$$F_{22}(k,t) = \frac{1}{\pi}\int_0^\infty u^2 g(r,t)\cos(kr)dr \tag{A.1.4}$$

with inverse transform,

$$u^2 f(r,t) = 2\int_0^\infty F_{11}(k,t)\cos(kr)dk \qquad (A.1.5)$$

$$u^2 g(r,t) = 2\int_0^\infty F_{22}(k,t)\cos(kr)dk \qquad (A.1.6)$$

Here $g$ and $f$ are the usual transverse and longitudinal correlation functions. Of course, $F_{11}(k,t)$ $F_{22}(k,t)$ are simply the one-dimensional energy spectra of $u_x(x,0,0)$ and $u_y(x,0,0)$.

By using the integral formula:

$$\int_0^\infty {}_1F_1(a;c;-t^2)\cos(2zt)dt = \sqrt{\frac{\pi}{2}}\frac{\Gamma(c)}{\Gamma(a)}z^{2a-1}e^{-z^2}U\left(c-\frac{1}{2},a+\frac{1}{2},z^2\right) \qquad (A.1.7)$$

where ${}_1F_1(a,c,z)$ is Kummer's hypergemometric function, and $U(a,c,z)$ is another function closely related to Kummer's function defined by

$$U(a,c,z) = \frac{\Gamma(1-c)}{\Gamma(1+a-c)}{}_1F_1(a,c,z) + \frac{\Gamma(c-1)}{\Gamma(a)}z^{1-c}\,{}_1F_1(1+a-c,2-c,z) \qquad (A.1.8)$$

where $\Gamma(z)$ is the usual Gamma function.

For the Sedov-type solution, according to formula (2.2), for isotropic turbulence, we have

$$F_{11}(k,t) = \frac{2\overline{u^2}}{\pi}\cdot\frac{l}{\sqrt{a_1}}\cdot\sqrt{\frac{\pi}{2}}\frac{\Gamma\left(\frac{5}{2}\right)}{\Gamma(\sigma)}\cdot\left(\frac{l}{\sqrt{a_1}}k\right)^{2\sigma-1}e^{-\frac{l^2}{a_1}k^2}U\left(2,\sigma+\frac{1}{2},\frac{l^2}{a_1}k^2\right) \qquad (A.1.9)$$

$$E(k,t) = \sqrt{\frac{8}{\pi a_1}}\cdot\left(\frac{2}{a_1}\right)^2\cdot\frac{\Gamma\left(\frac{5}{2}\right)}{\Gamma(\sigma)}\cdot(bl)\cdot(kl)^4\cdot e^{-\frac{l^2}{a_1}k^2}\cdot U\left(\frac{5}{2}-\sigma,\frac{7}{2}-\sigma,\frac{l^2}{a_1}k^2\right) \qquad (A.1.10)$$

This is the exact result of the spectra based on the Sedov-type solution. In further consideration, we want to know it is probably fair to say: Is this a satisfactory closure scheme which encompasses both the large and the small scales?

By using the definition of $U(a,c,z)$, we also have[6]

---

[6] This mathematical manipulation lies in the fact that the expression equ.(2.10) is not well defined since it has a pole $\Gamma(0)$, according to the definition (4.7), one can attempt to define new expression of the function by using recurrence relations to elimination this term.

$$U\left(\frac{5}{2}-\sigma, \frac{7}{2}-\sigma, z\right) = U\left(\frac{5}{2}-\sigma, \frac{5}{2}-\sigma+1, z\right) \tag{A.1.11}$$

We also note that

$$(a+z)U(a,c,z) + a(c-a-1)U(a+1,c,z) - zU(a,c+1,z) = 0 \tag{A.1.12}$$

$$U(a,c+1,z) = \left(1+\frac{a}{z}\right)U(a,c,z) + a(c-a-1)\cdot\frac{1}{z}\cdot U(a+1,c,z) \tag{A.1.13}$$

The substituting yields

$$E(k,t) = \sqrt{\frac{8}{\pi a_1}} \cdot \left(\frac{2}{a_1}\right)^2 \cdot \frac{\Gamma\left(\frac{5}{2}\right)}{\Gamma(\sigma)} \cdot (bl) \cdot (kl)^4 \cdot e^{-\frac{l^2}{a_1}k^2} \times$$

$$\left\{\left[1+\frac{\frac{5}{2}-\sigma}{\frac{(kl)^2}{a_1}}\right]U\left(\frac{5}{2}-\sigma,\frac{5}{2}-\sigma,\frac{l^2}{a_1}k^2\right) - \frac{\frac{5}{2}-\sigma}{\frac{(kl)^2}{a_1}}\cdot U\left(\frac{7}{2}-\sigma,\frac{5}{2}-\sigma,\frac{l^2}{a_1}k^2\right)\right\} \tag{A.1.14}$$

Based on the above formula, it is easy to handle the asymptotic analysis on the behavior of the energy spectra .

**A.2 The power-law spectra**

Sixty years ago, .A.N.Kolmogorov (1941) proposed an elegant theory of the universal statistical properties of small-scale eddies in high Reynolds number turbulent flows. Kolmogorov's theory (hereafter referred to as K41) is still the basis for nearly all work on the statistical theory of turbulence. By using physically motivated and dimensional arguments, spectrum estimates of some velocity moments have been obtained. Of course those arguments had not referred explicitly to the equations governing turbulent flows; a notable example is original "derivation" of the energy spectrum of full developed homogeneous isotropic turbulence by Kolmogorov. Curiosly enough, little attention has been paid to the possibility of eliciting physically meaningful result directly from the properties of the statistics governing turbulence. It is worth stressing that K41 makes no direct connection to the Navier-Stokes equations, furthermore, K41 is also not correct in detail (Sreenivassan[7], 1999).

The main task in connection with the theory of isotropic turbulence at present seems to be the prediction or explanation of this power law. This is still the main challenge for isotropic turbulence theory. This section devotes to this important issue. We will see that the power-law is one asymptotic state, derived naturally from the asymptotic expansion of the general turbulent spectra given above.

We remember that, the turbulent spectrum has been rewritten as eq. (A.1.10). It is natural to see what will follow if we adopt the asymptotic expansion in the case of the large argument. The

---

[7] K.R.Sreenivassan, Fluid turbulence. Reivew of Modern Physics, vol.71, no.2, 1999.

details have been listed from the mathematical point of view. Furthermore, we also have known that:

For the asymptotic expansions of the function $U(a,c,z)$, as $z \to \infty$, we have

$$U\left(\frac{5}{2}-\sigma, \frac{5}{2}-\sigma, \frac{l^2}{a_1}k^2\right) = \sum_{m=0}^{N} A_m \left(\frac{l^2}{a_1}k^2\right)^{-m-\left(\frac{5}{2}-\sigma\right)} \quad (A.2.1)$$

$$U\left(\frac{7}{2}-\sigma, \frac{5}{2}-\sigma, \frac{l^2}{a_1}k^2\right) = \sum_{m=0}^{N} B_m \left(\frac{l^2}{a_1}k^2\right)^{-m-\left(\frac{7}{2}-\sigma\right)} \quad (A.2.2)$$

where

$$A_m = (-1)^m \cdot \frac{\left(\frac{5}{2}-\sigma\right)_m \cdot (1)_m}{m!} \quad (A.2.3)$$

$$B_m = (-1)^m \cdot \frac{\left(\frac{7}{2}-\sigma\right)_m \cdot (2)_m}{m!} \quad (A.2.4)$$

Performing some mechanical calculations, we have

$$E(k,t) = \sum_{m=0}^{M} \sqrt{\frac{8}{\pi a_1}} \cdot \left(\frac{2}{a_1}\right)^2 \cdot \frac{\Gamma\left(\frac{5}{2}\right)}{\Gamma(\sigma)} \cdot \left(bl^{2(\sigma-m)-2}\right) \cdot e^{-\frac{l^2}{a_1}k^2} \times$$
$$\left\{\left[\frac{\frac{5}{2}-\sigma}{\frac{1}{a_1}}\right] \cdot A_m \cdot \left(\frac{1}{a_1}\right)^{(\sigma-m)-\frac{5}{2}}\right\} \cdot k^{2(\sigma-m)-3} \quad (A.2.5)$$

It should not be forgotten, however, that the entire discussion above is restricted to recover the power law spectra. While it is naturally to set that:

$$2(\sigma-m)-3 = -\frac{5}{3} \quad (A.2.6)$$

Hence,

$$\sigma_m = \frac{2}{3}+m \quad (A.2.7)$$

where $m = 0,1,2,....,M+1$.

The corresponding energy spectrum takes the form,

$$E(k,t) = C_L \cdot J(t) \cdot k^{-\frac{5}{3}} \quad (A.2.8)$$

where[8]

$$C_L = \sum_{m=0}^{M} \sqrt{\frac{8}{\pi a_1} \cdot \left(\frac{2}{a_1}\right)^2 \cdot \frac{\Gamma\left(\frac{5}{2}\right)}{\Gamma(\sigma)} \left\{ \left[\frac{\frac{5}{2}-\sigma}{\frac{1}{a_1}}\right] \cdot A_m \cdot \left(\frac{1}{a_1}\right)^{-\frac{11}{6}} \cdot \right\}} > 0 \cdot \quad (A.2.9)$$

$$J(t) = bl^{-\frac{2}{3}} \quad (A.2.10)$$

Now it is time we learn more things from K41. For small scales much larger than the Kolmogorov scale, one recovers the inertial range expression:

$$E_K(k) = K_0 \cdot \varepsilon^{\frac{2}{3}} \cdot k^{-\frac{5}{3}} \quad (A.2.11)$$

where $K_0$ is the Kolmogorov canstant.

It is important to note that in the present framework,

$$\varepsilon = \varepsilon_0 \left(bl^{-2}\right) \quad (A.2.12)$$

After a little work, one finds

$$E(k,t) = C_L \cdot \left(bl^2\right)^{\frac{1}{3}} \cdot \varepsilon^{\frac{2}{3}} \cdot k^{-\frac{5}{3}} \quad (A.2.13)$$

This is the final expression of the power law spectrum based on the exact solution. The connection of this conclusion with Kolmogorov's theory in inertial range is an interesting question. So, the main effects are readily recovered in our analysis.

In comparison, it seems that the conclusion drawn from this section could be listed as following:

[1] It is natural to come to the K41-like power-law spectra, if we adopt the asymptotic expansion in the case of the large argument and set $\sigma_m = \frac{2}{3} + m$; $m = 0,1,2,...,M$;

[2] It is important to note that the "constants" involved in these decay laws are to be dependent of time. This feature represents a crucial departure from K41.

## APPENDIX B: Equivalence of quadratic polynomials

We have followed Devaney's book for a definition of topologically conjugacy.

Let $X$ and $Y$ be two subsets of the real line and let $f$ and $g$ are said to be topologically conjugate provided $f$ and $g$ are continuous and there is a homeomorphism

$$h: X \rightarrow Y \quad (B.1)$$

---

[8] One can easily confirm that $C_L > 0$.

such that the functional equation

$$h(f(x)) = g(h(x)) \tag{B.2}$$

hold for all $x \in X$.

A mapping $h$ is said to be a homeomorphism provided $h$ is continuous, one-to-one and onto, and the inverse mapping $h^{-1}$ is also continuous.

We show that for any polynomial of second degree like

$$f(x) = \alpha x^2 + \beta x + \gamma \tag{B.3}$$

there is a homeomorphism $h$ such that

$$h(f(x)) = g(h(x)) \tag{B.4}$$

for all $x \in R.$
where

$$g(x) = x^2 + c. \tag{B.5}$$

In fact, $h$ can be chosen as an affine linear mapping

$$h(x) = mx + n. \tag{B.6}$$

It is easy to verify that

$$m = \alpha \tag{B.7}$$

$$n = \frac{1}{2}\beta \tag{B.8}$$

$$c = \alpha\gamma + \frac{\beta}{2}\left(1 - \frac{\beta}{2}\right). \tag{B.9}$$

Let us derive the coefficients for $h$ from the assumption that (B.6), and that $h$ solves the functional equation (B.4). Using the explicit forms of $f$ and $g$, and $h$ this yields

$$\begin{aligned}h(f(x)) &= m(\alpha x^2 + \beta x + \gamma) + n \\ &= m\alpha x^2 + m\beta x + (m\gamma + n)\end{aligned} \tag{B.10}$$

$$\begin{aligned}g(h(x)) &= (mx + n)^2 + c \\ &= m^2 x^2 + 2mnx + (n^2 + c)\end{aligned} \tag{B.11}$$

Comparing coefficients this gives

$$m = \alpha \tag{B.12}$$

$$n = \frac{\beta}{2} \tag{B.13}$$

$$c = \alpha\gamma + \frac{\beta}{2}\left(1 - \frac{\beta}{2}\right). \tag{B.14}$$

For isotropic turbulence problem we explored so far, we have

$$f(x) = -2Ax^2 + x + \frac{1}{2}B \tag{B.15}$$

The coefficients for $h$ read

$$m = -2A \tag{B.16}$$

$$n = \frac{1}{2}. \tag{B.17}$$

The final expression for the bifurcaion parameter $c$ is

$$\begin{aligned} c &= \alpha\gamma + \frac{\beta}{2}\left(1 - \frac{\beta}{2}\right) \\ &= (-2A) \times \left(\frac{1}{2}B\right) + \frac{1}{2} \times \left(1 - \frac{1}{2}\right). \\ &= \frac{1}{4} - AB \end{aligned} \tag{B.18}$$

24